\documentclass[twocolumn,showpacs,amsmath,amssymb, nobibnotes, aps, prl,showkeys]{revtex4}
\usepackage{graphicx}% Include figure files
\usepackage{dcolumn}% Align table columns on decimal point
\usepackage{bm}% bold math
\usepackage{docs}
\usepackage{boldline,multirow}
\usepackage[dvipsnames]{xcolor}
\expandafter\ifx\csname package@font\endcsname\relax\else
 \expandafter\expandafter
 \expandafter\usepackage
 \expandafter\expandafter
 \expandafter{\csname package@font\endcsname}%
\fi
\newcommand{\be}{\begin{equation}}
\newcommand{\ee}{\end{equation}}
\newcommand{\bea}{\begin{eqnarray}}
\newcommand{\eea}{\end{eqnarray}}

\begin{document}
\title{A pure hadronic model description of the observed neutrino emission from the tidal disruption event AT2019dsg}
\author{Prabir Banik$^{1}$\thanks{Email address: pbanik74@yahoo.com}}
\thanks{pbanik74@yahoo.com}
\author{Arunava Bhadra$^{2}$\thanks{Email address: aru\_bhadra@yahoo.com}}
\thanks{aru\_bhadra@yahoo.com}
\affiliation{ $^{1}$Department of Physics, $\&$ Center for Astroparticle Physics $\&$ Space Science, Bose Institute, EN-80, Sector-5, Bidhan Nagar,  Kolkata-700091, India  \\
$^{2}$High Energy $\&$ Cosmic Ray Research Center, University of North Bengal, Siliguri 734013, India}

%\affiliation{ $^{1}$High Energy $\&$ Cosmic Ray Research Centre, University of North Bengal, Siliguri, West Bengal, India 734013}
\begin{abstract}
Recently, the IceCube Neutrino Observatory has detected the neutrino event IceCube-170922A from the radio-emitting tidal disruption event (TDE) named AT2019dsg, indicating to be one of the most likely sources of high-energy cosmic rays. So far, the photo-hadronic interaction is considered in the literature to interpret neutrino emission from AT2019dsg. Here, we examine whether the IceCube-170922A along with the broadband electromagnetic emission from the source can also be described by a pure hadronic emission employing the proton blazar inspired (PBI) model, which takes into account the non-relativistic protons that emerge under the charge neutrality situation of the blazar jet and thus offers sufficient target matter for $pp$ interactions with shock-accelerated protons. Our findings show that the PBI model is able to consistently describe the IceCube observations on AT2019dsg and the broadband spectrum of the source without exceeding the observed X-ray and gamma-ray flux upper limits imposed by the XMM-Newton and Fermi-LAT telescopes.

\end{abstract}

\maketitle

\section{Introduction}
\label{intro}
On October 1, 2019, the IceCube Neutrino Observatory announced the detection of an intriguing high-energy muon-neutrino event IC191001A of energy $\sim 0.2$ PeV with a 56.5\% chance of being a genuine astrophysical neutrino \cite{Stein19}. The neutrino event is found to be associated with a radio-emitting tidal disruption event (TDE) designated AT2019dsg as identified by the Zwicky Transient Facility (ZTF) with a chance likelihood of detection of roughly 0.2 percent to 0.5 percent \cite{Stein20}. It is the first time a high-energy neutrino event has been correlated to a radio-emitting TDE \cite{Stein20}.

A TDE is an astrophysical phenomenon that occurs when a star is tidally disrupted by a supermassive black hole (SMBH) while passing close to the SMBH. It is generally believed that about half of the stellar debris can be accreted into the black hole and forms a transient accretion disk around the SMBH, resulting luminous flare \cite{Winter20}. The majority population of TDEs have a thermal spectrum in optical, ultraviolet, and x-ray bands. A small fraction of the bulk population of TDEs can launch a relativistic jet due to very high (super-Eddington) mass accretion rates \cite{Hills75,Rees88,Lacy82,Phinney89}, generating non-thermal radio emissions. Theoretically, jetted TDEs with non-thermal radio emission have been proposed as potential sources of ultrahigh-energy cosmic rays \cite{Farrar09,Farrar14}. The subsequent production of high energy gamma-rays in TDE environments have been historically explained via the lepto-hadronic ($p\gamma$) or pure hadronic ($pp$) interactions \cite{Wang11,Wang16,Dai17,Senno17,Lunardini17,Gupin18,Biehl18} since the discovery of the first jetted TDE Swift J1644+57 \cite{Burrows11}. Recent stacking searches suggest that the contribution of jetted and non-jetted TDEs to the diffuse extragalactic neutrino flux is limited to roughly 1\% and 26\%, respectively \cite{Stein20}.

The TDE AT2019dsg was identified by the ZTF survey on April 9, 2019 in the optical-UV band \cite{Nordin19}, which may be described by a blackbody photosphere with a near-constant temperature of $10^{4.59\pm 0.02}$ K and a photosphere radius of $10^{14.59\pm 0.03}$ cm \cite{van20}. The optical-UV band light-curve shows a peak luminosity of $10^{44.54\pm 0.08}$ erg/s which has reached a plateau of luminosity about $3\times 10^{43}$ erg/s at the arrival time of the neutrino event which is 150 days after the luminosity peak \cite{Stein20,Winter20}. The temperature of the object is in the top 5\% and its peak luminosity is in the top 10\% of the 40 known optical TDEs to date \cite{van20}. After 17 days from the optical-UV peak of AT2019dsg, the TDE was also detected in X-rays with a luminosity of about $2.5 \times 10^{43}$ erg/s in energy range of $0.3 - 8$ keV \cite{Winter20}. The observed X-ray spectrum is compatible with thermal emission and well described by a blackbody emission from a hot accretion disk of temperature $\sim 10^{5.9}$ K. This X-ray flux decreased very rapidly with time and this could be due to cooling of the newly-formed TDE accretion disk or rising X-ray obscuration \cite{Stein20}. The Fermi Large Area Telescope (Fermi-LAT) saw no significant signal from AT2019dsg and reported a gamma-ray flux upper limit of $10^{-12} - 10^{-11}$ erg cm$^{-2}$ s$^{-1}$ in the energy range of $0.1-800$ GeV averaged over 230 days after the TDE was identified \cite{Stein20}. About the localization of IC191001A, the HAWC observatory also investigated for transient gamma-ray emission on short timescales and set a upper limit of $E^2 dN/dE = 3.51 \times 10^{-13} (E/TeV)^{-0.3}$ TeV cm$^{-2}$ s$^{-1}$ with 95\% confidence for their most significant position, in the energy range 300 GeV to 100 TeV \cite{Ayala19}.

Recently, Winter and Lunardini (2020) \cite{Winter20} proposed that the TDE AT2019dsg features a relativistic jet that radiates neutrino and electromagnetic (EM) radiation along the line of sight. They suggested that a percentage of the X-ray photons are backscattered by the ionised electrons in the surrounding outflow, producing the target photon field for the generation of neutrinos through $p\gamma$ interactions \cite{Winter20}. According to Liu et al. (2020) \cite{Liu20}, the neutrino event and EM radiation of TDE AT2019dsg may be interpreted self-consistently in the context of an off-axis jet, in which relativistic protons are accelerated in the jet launched by the TDE and interact with its intense optical/UV radiation.  The quick decline of observed X-ray brightness with time was explained by the absorption of X-ray and gamma-ray photons by diverse processes by the high gas column density of the TDE outflow along the line of sight \cite{Liu20}. The TDE's outflow wind exhibits significantly varied density and velocity profiles at different inclination degrees, according to a three-dimensional completely general relativistic radiation magnetohydrodynamics simulation done in Ref.~\cite{Dai18}. In areas of decreasing density (away from the plane of the accretion disc and closer to the jet), the velocity profile of the gas revealed increasingly rapid outflows, with speeds exceeding $0.1c$ or perhaps $0.5c$ \cite{Winter20}. However, because no neutrino event was identified after the identification of IC191001A, the intrinsic X-ray and gamma-ray flux is expected to diminish over time. Murase et al. (2020) \cite{Murase20} suggested non-jetted scenarios, such as hot coronae surrounding an accretion disc, to explain the observed high-energy neutrino event and electromagnetic radiations from the TDE AT2019dsg. The broadband EM spectrum from radio to TeV energy has yet to be examined, which is important for understanding the source's true physical circumstances (e.g. magnetic field, Doppler factor, jet viewing angle, and other). 

Here we would like to show that the observed neutrino along with the multi-wavelength EM emission from the TDE AT2019dsg can well be interpreted as a purely hadronic emission assuming that the association of the observed neutrino events with the corresponding TDE is genuine. In this regard, we would like to exploit the main essence of the proton blazar inspired model \cite{Banik19,Banik20} at the jet of the TDE AT2019dsg to explain the observed multi-wavelength electromagnetic spectrum as well as the neutrino event from the source. Since TDE leads to formation of relativistic jets, the proton blazar model should be applicable for TDEs. Note that the optical polarimetry results of AT2019dsg are compatible with the existence of a spectral component that could be attributed to a jet, although such a structure was not confirmed \cite{Lee20}. 

The proton blazar inspired model can consistently explain the observed high-energy gamma-rays and neutrino signal from all the Icecube blazars during the flaring stage. Basically, the model explains the high-energy gamma-ray and neutrino observations based on interactions of shock accelerated protons in AGN jet with the cold protons present in the system ($pp$ interaction). In the framework of the stated model, the detected lower-energy bump of EM SED from the jet of a blazar can be well interpreted with the synchrotron radiation of relativistic electrons present in jet plasma, whereas the cold (nonrelativistic) proton density that arose from the charge neutrality condition can provide sufficient target matter (proton) for the production of high-energy gamma rays and neutrinos via the $pp$ interaction. It is worthwhile to mention that the radio observations manifest that AT2019dsg be a member of a rare type of TDEs which exhibit non-thermal radio emission produced due to synchrotron emission of accelerated electrons \cite{Stein20}.

The following is the structure of the paper: The approach for evaluating gamma-ray and neutrino fluxes in the framework of a proton blazar inspired model is described in the next section. Section III displays numerical estimates of the fluxes of generated multi-wavelength electromagnetic spectrum and high-energy neutrinos from the TDE AT2019dsg. The findings of the current work will be described in Sect. IV, and finally we will conclude in the same section. 

\section{The proton blazar inspired model in the context of TDE}
In a tidal disruption event, the companion star comes too close to a supermassive black hole and is finally torn apart by the strong gravitational (tidal) force of the black hole. Approximately half of the disrupted star material gets unbound from the SMBH, while the remaining plasma circularises to create an accretion disk. According to the Blandford-Znajek process \cite{Blandford77}, a weak initial magnetic field in the accretion disc combined with a strong black hole spin might result in the formation of a jet where particles (protons/nuclei, electrons) may be accelerated by internal shocks to ultra-high energies. This idea is supported by numerical simulations of TDEs based on general relativistic radiation magnetohydrodynamics (e.g. unified model by Dai et al. 2018 \cite{Dai18})

The overall jet composition of AGN/TDE remains an unknown issue so far. Almost all hadronic models of AGN/TDE jet usually assume that all the electrons and protons in a concerned system experience diffusive shock (Fermi) acceleration. In principle, cold (non-relativistic) protons resulting from charge neutrality are expected to occur in jets. The presence of cold protons in the blazar jet is indicated by a hybrid simulation study that showed that only a small percentage of the system's protons (about 4\%) are accelerated to non-thermal energy at diffusive shocks \cite{Caprioli15}. Meanwhile, the observationally acceptable range of the fraction of electrons population ($\chi_e$) that participates in diffusive shock (Fermi) acceleration is $m_e/m_p \le \chi_e \le 1$, but the exact fraction $\chi_e$ cannot be evaluated with current observations, as Eichler and Waxman pointed out in the context of Gamma-Ray Bursts \cite{Eichler05}. The adjustable parameters of the model include the ratio of the number of relativistic protons to electrons, the maximum energies achieved by protons/electrons throughout the acceleration process, the slope of their energy spectrum, and the luminosities of electrons and protons. A thorough computational approach for electromagnetic and neutrino spectra at the Earth from a blazar in the framework of a proton blazar inspired model is given in \cite{Banik19,Banik20}.

In the purview of the present model, the responsible region at the jet of a TDE for the non-thermal emission is a spherical blob of size $R_b'$ (primed variables for jet frame) and contains a tangled magnetic field of strength $B'$. The bulk Lorentz factor of the moving blob is $\Gamma_j = 1/\sqrt{1-\beta_j^2}$ and the corresponding Doppler factor of the blob is $\delta_{j} = \Gamma_j^{-1}(1-\beta_j\cos\theta)^{-1}$ where $\theta$ is the angle between the line of view and the jet axis \cite{Petropoulou15}. Both electrons and protons can be assumed to be co-accelerated in the jet in diffusive shock acceleration by parallel collision-less shock.

In the proton blazar inspired model framework, we assume a broken power-law energy distribution of shock accelerated electrons having spectral indices $\alpha_1$ and $\alpha_2$ before and after the spectral break at Lorentz factor $\gamma_b'$ as shown below \cite{Katarzynski01,Banik19}
\begin{eqnarray}
N_e'(\gamma_e') = K_e \gamma_e'^{-\alpha_1} \hspace{1.5cm} \mbox{if}\hspace{0.6cm} \gamma_{e,min}' \le \gamma_e' \le \gamma_b' \nonumber \\
         = K_e \gamma_b'^{\alpha_2-\alpha_1} \gamma_e'^{-\alpha_2} \hspace{0.35cm} \mbox{if}\hspace{0.46cm} \gamma_b' <\gamma_e' \le \gamma_{e,max}'\;
\label{Eq:1}
\end{eqnarray}
where $\gamma_e' = E_e'/m_e c^2$ is the Lorentz factor of electrons of energy $E_e'$ and %\textcolor{red}{
$K_e$ is the normalization constant that can be evaluated as (e.g. \cite{Bottcher13})
\begin{equation}
L_e = \Gamma_j^2 \pi R_b'^2 \beta_j c \int_{\gamma_{e,min}'}^{\gamma'_{e,max}}m_e c^2\gamma_e' N_e(\gamma_e') d\gamma_e'
\label{Eq:2}
\end{equation}
where $L_e$ represents the kinetic power of accelerated electrons in lab frame. If the spectral break in the electron distribution is a radiative (synchrotron+inverse Compton) cooling break in a uniform magnetic field within the emission region, the electron distribution breaks in its index by one power (i.e, $\Delta \alpha = \alpha_2-\alpha_1 \approx 1$) above the break Lorentz factor $\gamma_b'$ \cite{Inoue96,Longair94}. For cooling break, the electron break Lorentz factor, which corresponds to the radiative (synchrotron+inverse Compton) cooling time in Thomson limit ($\tau_{cool} \approx \frac{3m c}{4\sigma_T(u_B+u_{soft})\sigma_T \gamma_e'}$) equals the adiabatic timescale ($t_{ad} = 2 R_b'/c$), can be estimated as ($\gamma_b' \approx \frac{3m c^2}{8\sigma_T(u_B+u_{soft})R_b'}$) (e.g. \cite{Inoue96}), where $u_{soft}$ represents the soft photon energy density. The maximum energy of electrons can be calculated by equating radiative cooling time with acceleration timescale of electrons as given in Ref.~\cite{Inoue96} and can be written as $\gamma_{e,max}' \approx \left[ \frac{9 e B'}{80 (u_B+u_{soft})\sigma_T\xi}\right]^{1/2}$, where $\xi$ is the efficiency of scattering for the particle acceleration, referred as the gyrofactor \cite{Inoue96}. If $n_{e,h}'= \int N_e'(\gamma_e') d\gamma_e'$ be the total number density of relativistic (`hot') electrons then the total number electrons number density including `hot' and non-relativistic (`cold') electrons will be $n_{e}' = n_{e,h}'/\chi_e$ \cite{Banik19,Banik20}.

The synchrotron radiation of primary accelerated electrons represents the low-energy component of the EM SED ranging from radio to x-ray energies produced from a blazar/TDE jet, which is evaluated here following the formulation given in B$\ddot{o}$ttcher et al. (2013) \cite{Bottcher13}. The inverse Compton (IC) scattering of primary accelerated electrons with seed synchrotron photons co-moving with the TDE jet (i.e., the SSC model) generally provides a significant contribution to the observed EM spectrum in the MeV to GeV energies, and can be evaluated using the expressions given in Refs.~\cite{Banik19,Blumenthal70,Inoue96}.

In the proton blazar inspired model, cosmic ray protons are thought to be accelerated in the same region of the TDE/blazar jet, and the production spectrum is expected to follow a power law 
\begin{equation}
 N_p'(\gamma'_p) =  K_p {\gamma'_p}^{-\alpha_p} .
%\label{Eq:4}
\end{equation}
where $\gamma_p' = E_p'/m_p c^2$ denotes the Lorentz factor of accelerated protons of energy $E_p'$, and $\alpha_p$ denotes the spectral index. Here, $K_p$ denotes the proportionality constant that can be calculated from the same expression as Eq.~\ref{Eq:2} but for protons and $L_p$ is the corresponding jet power in relativistic protons. The energy density of relativistic protons is $u_p' = \int m_p c^2\gamma_p' N_p'(\gamma_p') d\gamma_p'$, while the number density of relativistic protons is $n_p' = \int N_p'(\gamma_p') d\gamma_p' $. When the shock-accelerated cosmic rays interact with the cold matter (protons) of density $n_{H} = (n_{e}'-n_p')$ in the blob of a jet, secondary particles (mainly pions) are produced \cite{Banik19}. Following Refs. \cite{Liu19,Anchordoqui07,Banik17a,Kelner06}, we have estimated the emissivity of secondary particles. The secondary neutral and charged pions finally decay to high-energy gamma-rays and neutrinos respectively. 

The produced TeV$-$PeV gamma-rays can be absorbed through internal gamma-gamma ($\gamma\gamma$) interactions while traveling through isotropic low-frequency synchrotron radiation produced by the relativistic electrons in the comoving jet \cite{Banik19}. The optical depth for the interaction as a function of photon energy $E_{\gamma}'( = m_e c^2 \epsilon_{\gamma}')$ in comoving jet frame is $\tau_{\gamma \gamma}(\epsilon_{\gamma}')$ which can be obtained following \cite{Aharonian08,Banik19}. The fraction of gamma-rays flux that will escape from the emission region of the jet after internal $\gamma\gamma$-absorption, can be expressed as $\left( \frac{1-e^{-\tau_{\gamma \gamma}}}{\tau_{\gamma \gamma}} \right)$ \cite{Bottcher13,Banik19}. 

The $\gamma\gamma$ interactions will create high-energy electron positron pairs in the emission region of the AGN/TDE jet. Also, the decay of $\pi^{\pm}$ mesons created in $pp$ interactions also produce high-energy electrons/positrons in the emission region. These injected secondary electrons/positrons carry the majority of the energy of the high-energy photon ($E_{\gamma}'$) the center-of-momentum energy greatly surpasses the rest energy of an electron (i.e, $\sqrt{E_{\gamma}' E_s'} \gg m_e c^2$, where $E_{s}'$ represents energy of the seed synchrotron photons present in the jet) \cite{Liu20}. These high energy electrons/positrons may cool down via synchrotron radiation or inverse-Compton interactions with low energy photons in the deep Klein-Nishina regime to produce again high gamma-rays \cite{Liu20}. The $\gamma\gamma$ annihilation of these new photons will again create electrons/positrons and this cyclic process will continue until the pair production opacity of the newly created photons goes below unity \cite{Liu20}.  Therefore, the injected secondary electrons/positrons will launch an EM cascades in the emission region of the jet via IC scattering and synchrotron radiation putting their energies in the keV-GeV energy range. We follow the self-consistent formalism of B$\ddot{o}$ttcher et al. (2013) \cite{Bottcher13} after the inclusion of the IC mechanism to determine the secondary-pair cascade processes. Once the equilibrium pair distribution is known, the associated stationary synchrotron and IC emission is evaluated as discussed above. Following Banik \& Bhadra (2019) \cite{Banik19}, we calculated the total gamma-ray emissivity $Q_{\gamma,esc}'(\epsilon_{\gamma}')$ that escaped from the blob of blazar/TDE jet. It encompasses all of the above-mentioned processes, such as synchrotron and IC radiation of accelerated electrons, gamma-rays produced by $pp$ interaction, and synchrotron photons of EM cascade electrons. In the case of a TDE like AT2019dsg, the keV flux upper limit observed by Swift-XRT and the GeV flux upper limit measured by Fermi-LAT might restrict the model. 

We calculated the optical depths for probable gamma-ray absorption through the Thomson scattering and BH processes in the ionised matter (protons or free electrons) present inside the jet's emission zone. Their values are much less than one, and hence their influence may be ignored. For the cross section of Thomson scattering and BH processes, we refer to Refs. \cite{Klein29,Chodorowski92,Liu19}. In the proton blazar inspired model framework, we assume entirely ionized matter inside the emission zone of the jet, hence we don't consider photoionization opacity.

The impact of extragalactic background (EBL) light on gamma-ray photon absorption was also presented here, and $\tau_{\gamma\gamma}^{EBL}(\epsilon_{\gamma},z)$ is the equivalent optical depth that may be derived using the Franceschini-Rodighiero-Vaccari (FRV) model \cite{Franceschini08,wabside}. Therefore, the measurable differential flux of gamma rays reaching Earth from a TDE jet may be expressed as \cite{Banik19}
\begin{eqnarray}
%\scriptsize
E_{\gamma}^2\frac{d\Phi_{\gamma}}{dE_{\gamma}} = \frac{V'\delta_{j}^4}{4\pi d_L^{2}}\frac{E_{\gamma}'^2}{m_e c^2} Q_{\gamma,esc}'(\epsilon_{\gamma}') e^{-\tau_{\gamma\gamma}^{EBL}}
\label{Eq:8}
\end{eqnarray}
where $V' = \frac{4}{3}\pi R_b'^3$ denotes the volume of the emission region, $E_{\gamma} = \delta_{j} E_{\gamma}'/(1+z) $ \cite{Atoyan03} represents photon energies in the observer, $d_L$ is the luminosity distance between the AGN and the Earth, and $z$ is the red shift parameter of the source. 

The charged pions generated by $pp-$interaction decay to high energy neutrinos, and the corresponding muon neutrino flux reaching the Earth may be represented as \cite{Banik19}
\begin{eqnarray}
%\scriptsize
E_{\nu}^2\frac{d\Phi_{\nu_{\mu}}}{dE_{\nu}} = \xi.\frac{V'\delta_{j}^4}{4\pi d_L^{2}} \frac{E_{\nu}'^2}{m_e c^2}Q_{\nu}'(\epsilon_{\nu}') 
\end{eqnarray}
where $\xi = 1/3$ owing to neutrino oscillation and $E_{\nu} = \delta_{j} E_{\nu}'/(1+z)$ \cite{Atoyan03} denotes the neutrino energies in the observer and co-moving jet frames, respectively. If the differential flux of muon neutrinos is known, the number of anticipated muon neutrino events in the IceCube detector in time $t$ may be calculated using the relation
\begin{eqnarray}
N_{\nu_{\mu}} = t \int_{E_{\nu,min}}^{E_{\nu,max}} A_{eff}(E_{\nu}). \frac{d\Phi_{\nu_{\mu}}}{dE_{\nu}} dE_{\nu}
\label{event}
\end{eqnarray}
where $A_{eff}$ is the IceCube detector effective area at the declination of the TDE in the sky \cite{Aartsen20}.

The entire jet power in the form of the magnetic field, relativistic electrons, cosmic rays, and cold matter with density $\rho'$ (including cold protons and electrons) is calculated using the relationship \cite{Banik19}
\begin{eqnarray}
L_{jet} = \Gamma_j^2 \beta_j c \pi R_b'^2\left[\rho'c^2(\Gamma_j-1)/\Gamma_j + u' +p'\right].
\label{jetpower}
\end{eqnarray}
where $u'$ (sum of $u_e'$, $u_p'$, and $u_B'$) represents energy density and $p' = u'/3$ (sum of $p_e'$, $p_p'$, and $p_B'$) denotes total pressure due to relativistic electrons, cosmic rays, and magnetic field in a comoving jet frame. At some point of time the cold protons, which will start losing energy in interaction with the environment, will fall back due to gravity but the higher energy particles will manage to escape as their (higher energy particles) jet power will remain below the Eddington luminosity. That is why we can separately estimated jet power of the relativistic particles (in jet frame) and radiations as
\begin{eqnarray} 
L^k_{jet} = \Gamma_j^2 \beta_j c \pi R_b'^2\left[u_e' + u_p' + u_B' \right]
\label{jetpowerk}
\end{eqnarray}
to compare with the system Eddington luminosity.

\section{Results}
The TDE AT2019dsg has a redshift of $z = 0.051$, implying a luminosity distance $D_L \approx 230$ Mpc from the Earth based on a consensus cosmology \cite{Nicholl19}. The black-hole mass of AT2019dsg's host galaxy was estimated to be $M_{bh} \simeq 3\times 10^{7} M_{\odot}$ based on the buldge mass estimation \cite{Stein20}. The corresponding Eddington luminosity of source is $L_{edd} = 1.3\times 10^{46}(M_{bh}/10^8M_{\odot}) = 3.9\times 10^{45}$ erg/s. The non-thermal radio emission and optical polarimetric studies indicate that the TDE AT2019dsg may have launched a relativistic jet, in which electrons and protons are accelerated to relativistic energy over the period of the neutrino event \cite{Stein20,Lee20}. Furthermore, the TDE AT2019dsg has the greatest sustained (over several days) X-ray brightness among the four TDEs discovered to have an X-ray counterpart out of the 17 TDEs in the ZTF sample \cite{Winter20}.

We consider that the relativistic jet has a bulk Lorentz factor of $\Gamma_j = 7$ and a Doppler boosting factor of $\delta_{j} = 13.7$ when seen at $\theta = 1^{\circ}$ angle. The adopted $\Gamma_j$ and $\delta_j$ parameters are similar to those used in Ref.~\cite{Winter20}. The size of the emission region within the jet can be assumed to be $R_b' =5\times 10^{14}$ cm  which is quite consistent with the size suggested from the variability, namely $R_b' \sim \delta_{j} c t_{ver}/(1+z) \simeq few\times 10^{14}$ cm considering the theoretical X-ray variability time scale of the jet $t_{ver} \simeq 630\; \frac{M_{bh}}{10^7 M_{\odot}}$ s \cite{Winter20}. The magnetic field within the emission region is considered to be $B' = 1.5$ G.

The lower energy part of the experimental EM SED data of the source, from radio to X-ray energy range, can be well explained by synchrotron emission of primary accelerated electrons obeying a broken power law with spectral indices $\alpha_1 = 2.1$ and $\alpha_2 = 3.1$ before and after the spectral break at the Lorentz factor $\gamma_b' = 7\times 10^{3}$. The spectral indices, and the electron break Lorentz factor under consideration are compatible with the spectral break of an equilibrium electron distribution caused by radiative cooling as discussed above \cite{Inoue96,Longair94}. The maximum Lorentz factor of accelerated electrons is found out to be $\gamma_{e,max}' = 10^6$ for the required  gyrofactor $\xi \approx 1000$. Note that gyrofactor can be as high as $\xi \approx 10^8$ as considered by Tanada et al. (2019) in their work \cite{Tanada19}. The kinematic power of relativistic electrons in the jet necessary to explain the observed data are determined to be $L_e = 3.5\times 10^{41}$ erg/s, respectively.  Furthermore, IC scattering of primary relativistic electrons with synchrotron photons co-moving with the AGN jet is shown to contribute significantly to the measured EM spectrum from the source at MeV to TeV energies.

In the proton blazar inspired model framework, the interactions of relativistic primary cosmic rays with the ambient cold protons in the blob produce high energy neutrinos and also contribute significantly to the EM SED above GeV energies. The necessary luminosity of injected accelerated primary protons is  $L_p = 3\times 10^{45}$ erg/s, with the best fitting spectral slope of $\alpha_p = - 1.9$ and the maximum achievable energy $E_{p,max}' \simeq 4\times 10^{15}$ eV. With the charge neutrality conditions, the cold proton number density in the jet is $1.2\times 10^7$ particles/cm$^3$, which offers enough targets for hadronuclear interactions with accelerated protons under the assumption of a low electron acceleration efficiency in the AGN jet of $\chi_e\approx 5 \times 10^{-3}$. The proton density of $10^7$ particles/cm$^3$ in blob is nearly the same to that of the TDE wind at the blob location at the age of the TDE (150 days) as demonstrated below. The numerical simulations of TDEs based on general relativistic radiation magnetohydrodynamics \cite{Dai18} suggest that the average mass accretion rate by the host SMBH (at near-peak times) can reach upto $\dot{M} \sim 10^2 L_{edd}$ (see also Ref.~\cite{De12,Guillochon13}). According to the simulation, a percentage of $\sim 20\% $, and $\sim 3\%$ of the total mass accretion rate go into the jet and the bolometric luminosity, respectively \cite{Winter20}. A remaining 20\% of the overall mass accretion rate powers the disk-driven wind \cite{Winter20}. Assuming a spherical distribution of wind-driven debris moving with velocity $v_w \sim (0.1-0.5)c$, the material density of the wind $r$ from the SMBH may be written as \cite{Murase20}
\begin{eqnarray}
\rho_w (r) = \frac{\dot{M}_w}{4\pi r^2 v_w}
\label{Eq:0}
\end{eqnarray}
where $\dot{M}_w = 20 L_{edd}/m_p c^2$ reflects the portion of the matter accretion rate that is attributed to the disk-driven wind. Wind debris with a velocity of $v_w \sim 0.3c$ can spread up to about $ 10^{17}$ cm in 150 days from the start of the event. However, we can consider that the emission region is located in the jet at a distance of $r = R_b'/\theta_j = 3.5 \times 10^{15})$ cm (as $R_b' = 5\times 10^{14}$ cm) from the black-hole where $\theta_j = 1/\Gamma$ represents the jet opening angle.  From Eq.~\ref{Eq:0}, one can estimate the matter density at the location of the blob in the wind to be $3 \times 10^7$ particles/cm$^3$. Note that the proton density of $\sim 10^7$ particles/cm$^3$ in blob corresponds to initial total proton number about $5\times 10^{51}$ and consequently the total mass of the blob will be nearly the same to that of the Earth which is very small in comparison to the mass of the parent system. 

\begin{figure}[t]
  \begin{center}
% \scalebox{2.5}{
  \includegraphics[width = 0.5\textwidth,height = 0.45\textwidth,angle=0]{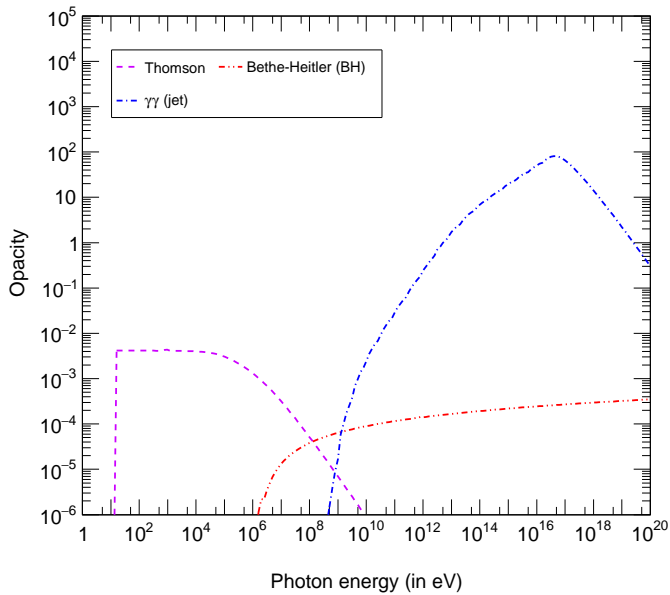}
\end{center}
%  \captionsetup{margin=50pt,font=small,labelfont=bf}
  \caption{Photon opacities calculated from various procedures. The pink small dashed, red dash-double-dotted and blue dash-single-dotted lines indicate the opacities of the Thomson scattering, BH process and $\gamma\gamma$ absorption for gamma rays, respectively.}
\label{Fig:1}
\end{figure}

\begin{figure*}[t]
  \begin{center}
% \scalebox{2.5}{
  \includegraphics[width = 1.0\textwidth,height = 0.45\textwidth,angle=0]{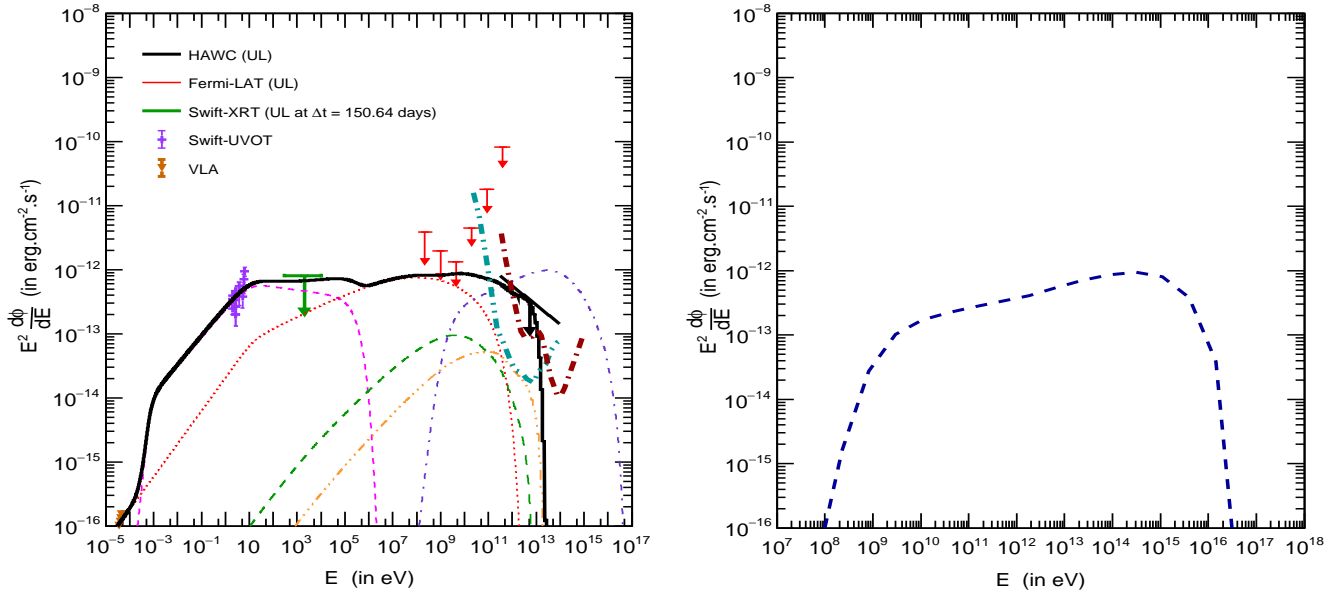}
\end{center}
%  \captionsetup{margin=50pt,font=small,labelfont=bf}
  \caption{Left: The estimated differential energy spectrum of the EM radiation from the emission region of the jet of TDE AT2019dsg. The EM spectra due to synchrotron emission and the inverse Compton emission by the relativistic electrons are shown by the magenta small-dashed and green long-dashed lines, respectively. The violet small dash-single-dotted line denotes the gamma-ray flux produced from pp interactions after $\gamma\gamma$ absorption. The red dotted and orange large-dashed triple-dotted lines respectively represent the flux for the synchrotron and the inverse Compton emission by electrons/positrons in the EM cascade after $\gamma\gamma$ absorption inside the emission zone of the jet. The black continuous line shows the estimated overall differential multiwavelength EM SED including the absorption in EBL. The detection sensitivity of the CTA detector for 1000 hours and the LHAASO detector for one year are represented by the cyan long dash-double-dotted and brown long-dash-single-dotted lines lines, respectively.  Right: The estimated neutrino flux reaching the Earth from the TDE AT2019dsg. }
\label{Fig:2}
\end{figure*}

As previously mentioned, the contributions to the observed EM spectrum owing to synchrotron and inverse Compton emission of stationary electron/positron pairs generated in an EM cascade triggered by protons are calculated. The produced gamma-rays are likely to be absorbed significantly in TeV$-$PeV energy band due to internal gamma-gamma ($\gamma\gamma$) interactions while traveling through isotropic low-frequency synchrotron radiation produced by the relativistic electrons in the comoving jet. The opacities of gamma-ray photons via internal $\gamma\gamma$ absorption in synchrotron photon field, Thomson scattering by free cold electrons and BH process by cold protons inside the emission zone of the jet are shown in Fig.~\ref{Fig:1}. It is found that Thomson scattering and BH process opacities are much less than one, and hence their influence may be ignored. The estimated differential gamma-ray spectra escaping from the emission area in the jet, as well as other satellite and ground-based data, are shown in the left panel of Fig.~\ref{Fig:2}. As seen from the image, the derived EM spectrum from the jet remains consistent (not exceeding) with the X-ray and gamma-ray flux limitations set by Swift-XRT (at 150.64 days following the discovery of the TDE) and the HAWC observatory, respectively. Note that here we did not consider the XMM-Newton X-ray flux limit which is given at 196.16 days following the discovery of the TDE, because of non-detection of any neutrino event during this time. Furthermore, the diminishing nature of the radio luminosity above 178 days after the discovery of the TDE indicates that the intrinsic X-ray and gamma-ray flux is likely falling over time, \cite{Stein20}, which explains the XMM-Newton X-ray flux limit. 

\begin{table}[t]
%  \begin{center}
    \caption{Model fitting parameters for the TDE AT2019dsg according to proton blazar model.}
    \label{table1}
%    \begin{tabular}{c|c}
    \begin{tabular}{ll}
 \hline\noalign{\smallskip}
 %     \toprule
      Parameters &  Values   \\ %\hline
\noalign{\smallskip}\hline\noalign{\smallskip}
       $z$                  &  $0.051$   \\
      $\Gamma_j$              &   $7$  \\
      $\theta$              &   $1^{\circ}$  \\
      $\delta_{j}$              &  $13.7$ \\
      $R_b'$  (in cm)            &  $5\times 10^{14}$ \\
       $B'$ (in G)          &  $1.5$  \\
       $\alpha_1$           &   $- 2.1$  \\ 
       $\alpha_2$           &   $- 3.1$  \\
       $\gamma_b'$          &  $7\times 10^{3}$ \\
       $\gamma_{e,min}'$    &  $1$ \\
       $\gamma_{e,max}'$    &  $10^{6}$ \\
       $L_e$ (in erg/s)    & $3.5\times 10^{41}$  \\ 
       $n_H$  (in cm$^{-3}$) &  $1.1\times 10^7$  \\
       $\alpha_p$           &   $- 1.9$  \\
       $E_{p,max}'$ (in eV) &  $4\times 10^{15}$  \\ 
       $L_p$ (in erg/s)    &  $3\times 10^{45}$  \\  %\\ %\hline    \hline
\noalign{\smallskip}\hline    
    \end{tabular}
%  \end{center}
\end{table}
The neutrinos produced in the emission zone of the jet will escape without any absorption in the TDE outflow because of their weakly interacting nature. The predicted differential neutrino flux reaching Earth is displayed in the right panel of the Fig.~\ref{Fig:2}. Using the Eq.~(\ref{event}), the expected muon neutrino event in IceCube detector from AT2019dsg is evaluated which turns out to be about $N_{\nu_{\mu}} = 0.45$ events in the energy range 1 TeV to 10 PeV in 154 days. The estimated number of neutrino events is conservative, since a probable contribution to the event rates from tau-neutrinos that produce muons with a branching ratio of 17.7 percent was not addressed \cite{Ansoldi18,Banik20}. The total muon like neutrino occurrences thus given as $N_{\mu}^{like} = N_{\mu} + 17.7\% \times N_{\mu}$ which becomes $0.53$ events.

The overall kinetic jet power in terms of relativistic electron, proton kinetic energy, and magnetic field is $3\times 10^{45}$ erg/s, while the total physical jet power is $2\times 10^{46}$ erg/s. Our estimated kinetic and physical jet power are $0.8$ and $5.1$ times greater than the source's Eddington luminosity, respectively, which are consistent with the supper-Eddington accretion rate predicted by Dai et al. (2018) in their unified model of TDEs \cite{Dai18}. A comparatively larger viewing angle of the jet corresponds to greater jet power.  

\begin{figure}[h]
  \begin{center}
% \scalebox{2.5}{
  \includegraphics[width = 0.5\textwidth,height = 0.45\textwidth,angle=0]{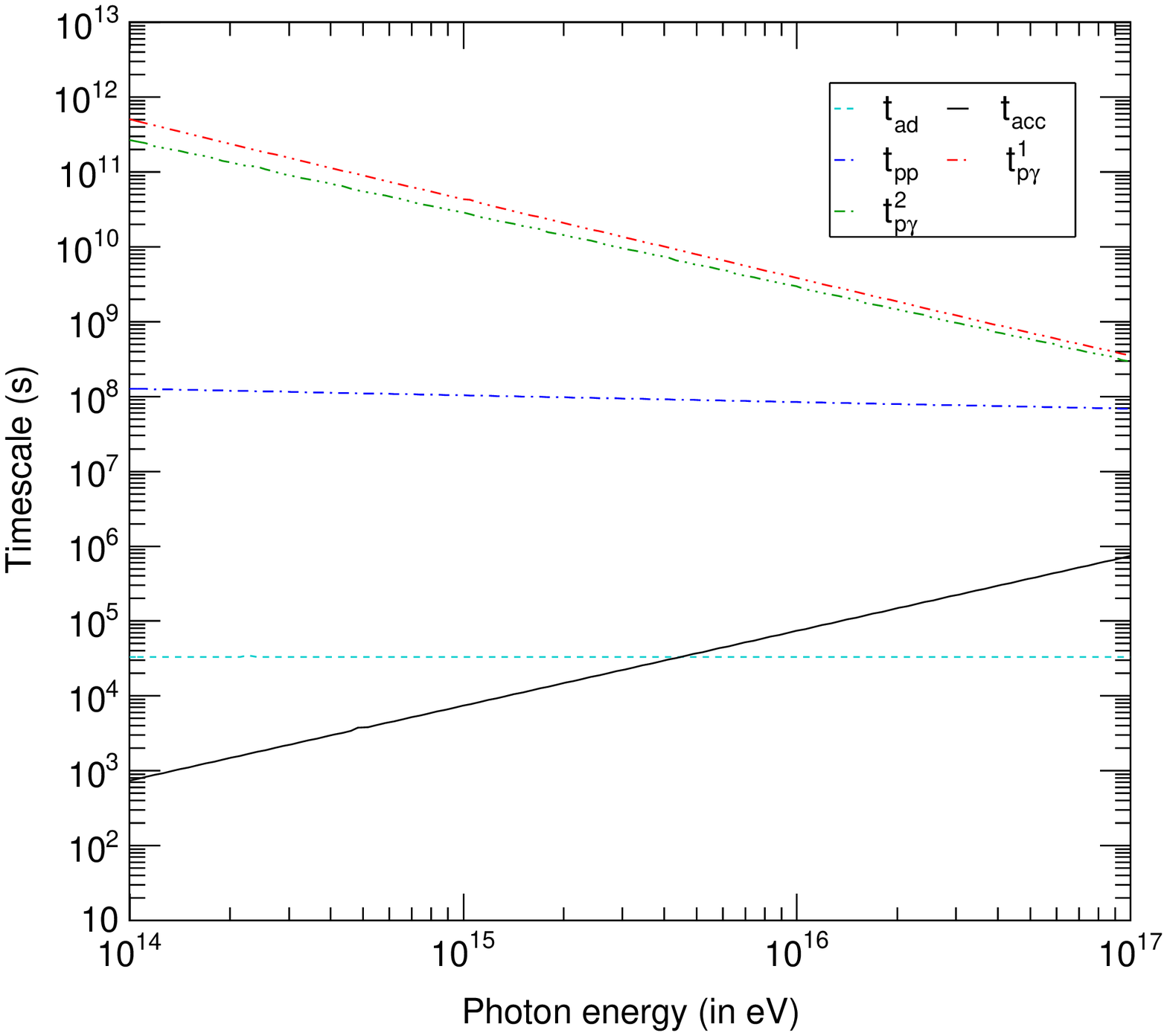}
\end{center}
%  \captionsetup{margin=50pt,font=small,labelfont=bf}
  \caption{The estimated relevant timescales for protons. %\textcolor{red}{
	The black continuous, cyan small-dashed, and blue dash-single-dotted lines represent the acceleration, adiabatic, and $pp$ interaction timescales respectively. The red dash-double-dotted and green dash-triple-dotted lines represent the $p\gamma$ interaction timescales of accelerated protons with synchrotron photons and total photon distribution (shown by the black continuous line in Fig.~\ref{Fig:2}) respectively. }  
	%}
\label{Fig:3}
\end{figure}

\section{Discussion and Conclusion}
We consider that protons are accelerated in the jet via diffusive shock (Fermi) acceleration mechanisms. The timescale of acceleration in the co-moving jet frame may be represented by a generic formula $t_{acc} = \frac{\eta E_p'}{ eB'c}$, where $B'$ is the magnetic field in the jet, $e$ is the electric charge, and $\eta $ is the acceleration efficiency \cite{He18}. In a co-moving jet frame, the maximum energy of the protons may be estimated as $E_{p,max}' \simeq 4\times 10^{15}$ eV (we set $\eta = 100$) by comparing the acceleration timescale with the adiabatic timescale $t_{ad} \simeq 2 R_b'/c$ sec \cite{Ansoldi18} as displayed in Fig.~\ref{Fig:3}. In the observer frame, the highest energy of the cosmic ray particle possible in the TDE jet is found out to be $5.5 \times 10^{16}$ eV.

The right production scenario of observed neutrinos and gamma rays from putative astrophysical sources is another important issue. The ultra high energy neutrinos are generally believed to be produced in the sources by either pure hadronic ($pp$) or photo-meson ($p\gamma$) interactions. We estimated the timescale of $p\gamma$ interactions ($t_{p\gamma}^1$) of primary relativistic protons with the synchrotron photons co-moving with the TDE jet following Refs.~\cite{Banik17a,Stecker73} and also the timescale of $p\gamma$ interactions ($t_{p\gamma}^2$) of primary relativistic protons with the estimated total photon distribution which includes all secondary photons (shown by the black continuous line in Fig. ~\ref{Fig:2}). Note that we did not consider the presence of external photons (e.g. disc radiation, Radiation from the X-ray corona etc.) within the emission region of the co-moving jet in the proton blazar inspired model framework. In case of pure hadronic ($pp$) interaction of relativistic protons with cold protons in the emission region of the jet as discussed above, the interaction timescale $t_{pp} = \frac{1}{k_{pp}\sigma_{pp} n_H c}$, where $k_{pp} = 0.45$ and $\sigma_{pp}$ represent the inelasticity \cite{Gaisser90} and the cross section \cite{Kelner06} respectively for $pp$ interaction. The aforementioned timescales of relativistic protons as functions of proton energy are displayed in Fig.~\ref{Fig:3}. The figure indicates that the pure hadronic ($pp$) mechanism dominates over the $p\gamma$ interaction mechanism in the adopted proton blazar inspired model framework and in the present case one can safely ignore the contribution from the $p\gamma$ interactions. 

In this work, we have shown that the association between IceCube-170922A and TDE AT2019dsg may be interpreted as pure hadronic emission employing the proton blazar inspired model after incorporating possible gamma-ray absorption in the dense outflowing wind matter around the jet of the TDE. 
The noteworthy point is that the proton blazar inspired model can consistently describe the neutrino emission from all the IceCube point sources till now.

We found that the future telescopes like CTA \cite{Ong17} and LHAASO \cite{Liu17}, which have higher sensitivity than the present generation gamma ray telescopes, should be able to detect the EM SED of the source in a wide energy band and hence can help to identify the true physical model that leads to its neutrino emission.

\section*{Acknowledgments}
The authors would like to thank an anonymous reviewer for useful comments that helped us to improve the manuscript. PB acknowledges financial assistance from the SERB (DST), Government of India, under the fellowship reference number PDF/2021/001514.

\end{document}